# Designing and Implementing Future Aerial Communication Networks

Sathyanarayanan Chandrasekharan, Karina Gomez, Akram Al-Hourani, Sithamparanathan Kandeepan,
RMIT University, Melbourne, Australia
Tinku Rasheed and Leonardo Goratti, Create-Net, Trento, Italy
Laurent Reynaud, Orange, Lannion, France
David Grace, University of York, United Kingdom
Isabelle Bucaille, Thales Communications & Security, Paris, France
Thomas Wirth, Heinrich Hertz Institute, Berlin, Germany
Sandy Allsopp, Allsopp Helikites Ltd, Damerham, England

*Abstract* — **Providing *"connectivity from the sky"* is the new innovative trend in wireless communications. High and low altitude platforms, drones, aircrafts and airships are being considered as the candidates for deploying wireless communications complementing the terrestrial communication infrastructure. In this article, we report the detailed account of the design and implementation challenges of an aerial network consisting of LTE-Advanced (LTE-A) base stations. In particular, we review achievements and innovations harnessed by an aerial network composed of Helikite platforms. Helikites can be raised in the sky to bring Internet access during special events and in the aftermath of an emergency. The trial phase of the system mounting LTE-A technology onboard Helikites to serve users on the ground showed not only to be very encouraging but that such a system could offer even a longer lasting solution provided that inefficiency in powering the radio frequency equipment in the Helikite can be overcome.**

*Index Terms*— **Low Altitude Platforms, Aerial Base Stations, LTE-A, Aerial-Terrestrial Communications, Aerial Networks.**

## I. INTRODUCTION

The advances in microelectronics have diminished the size and weight of the wireless network equipment allowing the exploration of new ways to deploy wireless infrastructure. In recent times there have been increasing interest in aerial communication networks as shown by research and industry efforts. Different use cases have been envisioned for aerial network deployment, including public safety, in order to provide coverage and capacity to personnel during emergency and temporary large-scale events, and Internet connectivity in emerging countries. Several projects have launched initiatives to study the possibility of using aerial platforms for providing wireless services. Moreover, Google and Facebook are investigating the prospect of using aerial platforms to bring Internet access in emerging countries.

A pioneering project called CAPANINA looked at both mechanically and electronically steerable antennas to deliver broadband wireless access using high altitude platforms [1]. The Google Loon experiment is an ambitious project intended to provide network coverage to rural and remote areas. In particular, the underlying technology is presented as part of Google's plans to fund and develop wireless networks in emerging markets. As far as the project Loon is concerned, a fleet of high-altitude balloons, operating at an altitude of about 20 km in the low stratosphere, will be coordinated to cover specific large geographical areas to offer users with wireless services, at best, similar bit rates as those of 3G [2]. Facebook is also working on ways to provide Internet to people from the sky exploring a variety of technologies including high-altitude long-endurance planes and satellites. In order to achieve its objectives, Facebook is creating partnerships with aerospace and communications technology experts, including NASA's Jet Propulsion Laboratory, Ames Research Center and Ascenta [3].

This article reports the outcomes of the ABSOLUTE project[1], which aimed to design and implement LTE-A aerial base stations (AeNB) using Low Altitude Platforms (LAP) to provide wireless coverage and capacity for public safety usage during and in the aftermath of large-scale unexpected and temporary events [4]. The main goal of the project was to design and validate the next-generation of aerial networks to provide a reliable communication network which could be rapidly rolled out and integrated with satellite and LTE-A terrestrial networks, and which is flexible, scalable and interoperable. LTE-A was selected as the candidate technology due to its higher performance and flexibility in comparison with other technologies such as Wi-Fi or WiMAX. LAPs were chosen instead of high altitude platforms due to the advantages they offer in terms of rapid deployment and lower implementation cost.

A LTE-A base station was mounted on an aerial platform and trialed to measure the performance of such future aerial networks. This article provides a detailed account of designing and implementing the next-generation of aerial networks to provide wireless services. We include a discussion on the available aerial platforms which could be used for the wireless provisioning in Section II. Aerial networks regulation aspects are summarized in Section III. While the details of the aerial network implementation are provided in Section IV. Then, communication aspects regarding main challenges and limitations of AeNB are discussed in Section V. Finally, conclusions and future paradigms are provided in Section VI.

Table 1: Aerial platform comparison based on capabilities for carrying wireless communication systems.

| AERIAL PLATFORM CAPABILITIES | DRONES | AIRCRAFT | AIRSHIP | TETHERED HELIKITE |
|---|---|---|---|---|
| | 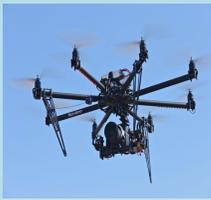 | 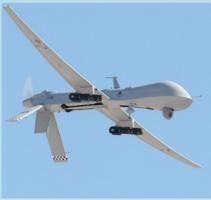 | 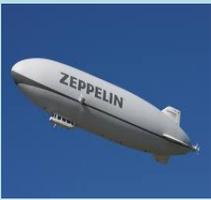 | 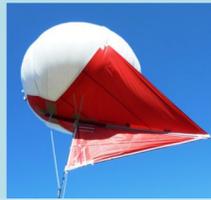 |
| High Payload (1-10Kg) | Depends on size | ✓ | ✓ | ✓ |
| Wide Area Coverage | ✓ | ✓ | ✓ | ✓ |
| Moving Coverage | ✓ | ✓ | ✓ | |
| Optimum Altitude | ✓ | ✓ | ✓ | ✓ |
| Extreme Duration | | ✓ | | |
| Ad-Hoc Network Friendly | ✓ | | ✓ | ✓ |
| Safe for Operators | ✓ | ✓ | | ✓ |
| Low Attrition Rate | | | | ✓ |
| Instant Deployment | ✓ | ✓ | | ✓ |
| Operation under Several Types of Weather Conditions | | ✓ | | ✓ |
| Deployment under Several Types of Weather Conditions | | ✓ | | ✓ |
| High Technology Security | | | | ✓ |
| Small & Easily Handled | ✓ | | | Depends on size |
| Single Person Deployment | ✓ | | | ✓ |
| Airborne Deployment | ✓ | ✓ | | |
| Air Traffic Friendly | Depends on the altitude | | | ✓ |
| Minimal Training | | | | ✓ |
| No Fuel Required | | Depends on the type | Depends on the type | ✓ |
| Good Antenna Placement | Depends on size | ✓ | ✓ | ✓ |
| Widely Available | ✓ | ✓ | | ✓ |
| Worldwide Operations | | ✓ | | ✓ |

## II. Aerial Platform For Wireless Services

In the ABSOLUTE project, the use of aerial platforms for providing wide-area wireless coverage was fundamental. The elevated look-angle provided by aerial platforms offers significant communication advantages compared to terrestrial equivalents. Moreover, it also offers the potential to deploy cameras or other sensors at the same time. Aerial platform based wireless communications are dependent upon a) the radio frequency equipment itself, and b) the physical characteristics of the platform. For such reasons, the capabilities of the most relevant aerial platforms available for the purposes of implementing aerial networks are summarized in Table 1.

### A. Drones Characteristics

Drones are a special type of unmanned aerial vehicles that are popular especially for remote sensing, photography and video surveillance [5]. Due to their relatively low capacity, both in terms of payload and autonomy, they are generally restricted to low or even very low altitudes (i.e. within a range of few hundred meters). Due to their small form factor, micro-drones can lift a very limited weight. Generally, the payload ranges from a few dozen of grams for the micro-drones to 5-7 kilograms for the larger drones. Due to their size, drones generally use lightweight lithium-ion batteries, powering the whole platform (propulsion, telemetry and payload included). Thus, the expected autonomy of drones is generally in the range of 10 to 40 minutes, depending mainly on the battery capacity, mission mobility pattern and payload weight. Drone's features and characteristics were not likely suitable for the ABSOLUTE scenarios where at least 10 kg payload (LTE equipment) carried at hundreds or thousands of meters is required.

### B. Aircraft Characteristics

One of the most widely used stratospheric unmanned aircraft is the Global Hawk. The Global Hawk was developed by Northrop Grumman for the US military, but is also being used for civilian use by NASA [6]. Powered by liquid fuel, it has significant payload capability. QinetiQ's Zephyr is a solar powered unmanned aircraft that is capable of remaining aloft for days. This aircraft is equipped with batteries that are charged during the day using solar energy, and then this stored energy is used during the night to allow it to remain airborne and stationary. Its payload capabilities are extremely limited, typically restricted to a maximum of 1 kg payload. The Ascenta-Hale is another solar powered unmanned aircraft capable of remaining aloft for 3 months or more carrying a payload of up to 25 kg. It is currently at the concept stage and is intended for both military and civilian applications [7]. This category of aerial platforms possesses favorable features such as low-power and energy-efficient lightweight structures with sufficient payload capacity, user-friendly interfaces which allow efficient trajectory management and positioning tools. However, the cost of the aircraft was the limiting characteristic for not choosing these aerial platforms in the context of the ABSOLUTE project.

*C. Airships Characteristics*

These types of aerial platforms which utilize lighter gas to float in air are classified as aerostatic platforms. Airships are much more flexible in terms of weight, size and power consumption of the payload, essentially only depending on the volume of the envelope (which can measure more than 100 m in length). However, the larger the volume, the bigger are the problems with keeping the airship stationary. Airships can be and have been designed for different altitudes. While commercial manned airships for cargo or passengers typically fly at low altitudes of approximately 200 m to save helium, unmanned airships have been designed to fly up to almost 30 km above the ground level. If keeping the airship stationary above the service area at selected operating altitude can be guaranteed with suitable electric motors and propellers, unmanned airships are capable of staying in the air for long periods of time, even years. The main drawback for the use of airships in disaster recovery scenario actually comes from their size, requiring high-strain envelope material, extensive ground operations center and appropriate ground facilities including hangars for storing and field for lifting and descending.

*D. Helikites Characteristics*

The name Helikite relates to the combination of a helium balloon and a kite to form a single, aerodynamically sound tethered aircraft, that exploits both wind and helium for its lift. The balloon is generally oblate-spheroid in shape [4]. The aerodynamic lift is essential to combat the wind and allows even small Helikites to fly at very high altitudes in high winds that push simple balloons to the ground. Helikites are very popular low altitude aerostatic platforms operable in several types of weather conditions. Thousands are operated worldwide, flown over land and sea by both civilians and the military. Helikites were chosen as the preferred aerial platform for the ABSOLUTE project due to the following characteristics:

▪ *Altitude:* Helikites utilize both wind lift and helium lift to enable high altitude flight in several types of weather conditions.
▪ *Payload:* Helikites can carry more payload than any other aerostat in high or low wind conditions.
▪ *Endurance and Cost:* Helikites need no electrical power to operate a ballonet and lose very little helium through their gas-tight inner balloon. Helikites are comparatively inexpensive to buy compared to other aerial platforms.
▪ *Regulations:* Helikites have very few legal obstacles compared to most other aerial platforms such as manned aircraft or drones.

III. REGULATIONS OF AERIAL NETWORKS

With the increasing popularity of aerial networks come many regulatory and legislative challenges, knowing that drones themselves are already creating a hot debate in some countries due to safety and privacy concerns. Aerial networks whether utilized for commercial coverage or emergency recovery operates under the civilian laws unless a severe disaster occurs which requires the intervention of the military. Thus, in the vast majority of cases aerial networks are subject to civilian regulations and licensing in order to guarantee the seamless deployment and operation in conjunction with other terrestrial wireless services, and in harmony with air traffic. The main legislative challenges can be categorized into two groups: aeronautical and radio regulations. There are different rules and regulations guarding aerial platforms depending on many factors such as:

▪ The category of the platform (Aircraft, Balloon, Airship, etc.)
▪ The control method of the platform (remotely piloted aircraft, tethered Helikites)
▪ The flying altitude, where it is usually allowed to fly below a certain altitude without a license, such as 120m in Australia
▪ The region of flying whether it is above an urbanized area, regional areas, or near to airports
▪ The situation of flying during an emergency, bushfire, or a normal operation, since some countries like Australia bans the flying of drones, model aircrafts or multi-rotor near bushfires, floods and traffic accidents.

Finally, well established regulations exist to assure flight safety and aeronautical frequency spectrum protection. Also, several new regulations in EU, USA and Australia were introduced regarding the radio control of aerial platforms. However, the main challenge resides in provisioning the wireless service itself and ensuring its minimal disturbance to existing terrestrial services. The International Telecommunication Union released several recommendations dealing primarily with HAPs such as M.1456 (05/00), M.1641 (06/03) and SF.1601-2 (02/07). More regulatory efforts are still required for better efficiency in exploiting spectral holes using cognitive radio techniques.

IV. AERIAL NETWORK IMPLEMENTATION

In this section, the details of an Aerial Base Station (AeNB) able to operate at 150 m altitude with 5 hours of autonomy are described. The AeNB is based on alternative network architecture for LTE deployment in which the majority of the base station equipment is contained in the base band unit (eNB-BB) and the radio frequency equipment in the Remote Radio Head (RRH) placed as close as possible to the antenna.

The RRH is connected to the eNB-BB via a fiber optic link reducing the coaxial feed line losses and providing a high level of flexibility in cell site construction. Thus, the AeNB is composed of both *i) the aerial segment* operating at varying altitudes in the air (RRH and antennas), and *ii) the terrestrial segment* operating stationary in the ground (eNB-BB plus distributed-Evolved Packet Core), both segments are linked using fiber optic link (see the Figure 1(a) ).

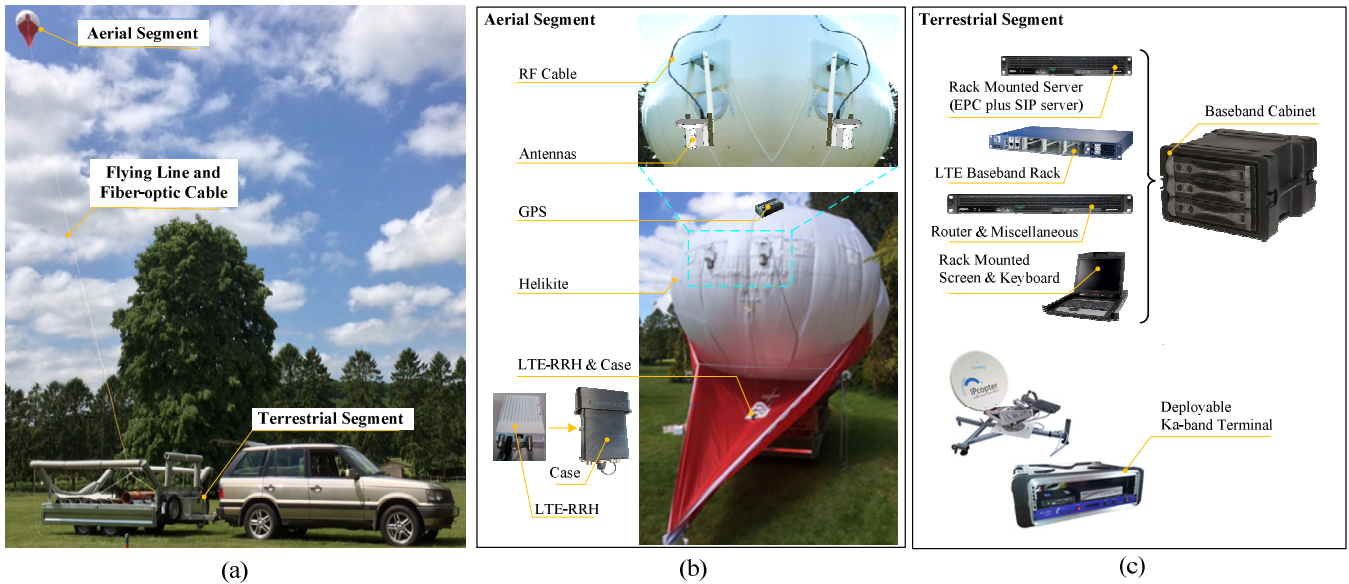

(a)  (b)  (c)

**Figure 1.** LTE-based aerial base station designed and implemented in ABSOLUTE project as combination of (a) aerial segment carrying RRH and antennas systems, and (b) terrestrial segments carrying eNB-BB, EPC and satellite systems. *Project consortium successfully demonstrated the usage of AeNB to the European Commission reviewers and end-user in Paris on September 30th 2015.*

### A. Aerial Segment

The aerial segment was the most challenging part during design and implementation of the AeNB, the Helikite being the crucial element, where the battery, antenna and RRH components are placed as shown in Figure 1(b).

#### a. Helikite Details

ABSOLUTE project chose a 34m³ desert star helikite, which is a special kite/balloon combination that uses both helium and wind for aerodynamic lift. The design of light and efficient RRH, waterproof suitcase, antennas and other equipment integrated in the Helikite is required due to helikite's payload limitations. The Helikite itself was tested in several types of weather conditions to ensure robustness and good stability, ease of set-up and handling and correct payload attachment webbing points. An automatic cut-down device (GPS integrated) is also installed as required by European regulations to ensure that the aerostat will bring itself down in the event it escapes its tether.

#### b. RRH Electronics, Case and Batteries

The RRH was updated to support a wider frequency range (up to 6 GHz) and to optimize the current consumption (1.7-1.8 A). The RRH platform is compact and has a lighter weight for compatibility with the aerial platform requirements. The RRH is composed of flexible software-defined radio (SDR) platform consisting of stacked digital interface card and radio frequency front-end, which allows commanding 2 radio frequency transceivers with 2-antenna duplex operation ranging from 3 MHz to 50 MHz radio frequency signal bandwidth. The RRH also supports cognitive extension functionality for dynamic spectrum allocation [8].

#### c. Antennas and Damped Pendulum Mount

Helix antennas radiation pattern has been shaped to illuminate the considered cell with a quasi-uniform power. Special attention has been paid to design very lightweight antennas. To achieve this, metalized foam was used to realize all antennas. Radiation and impedance results of Helix antenna are shown in Figure 2.

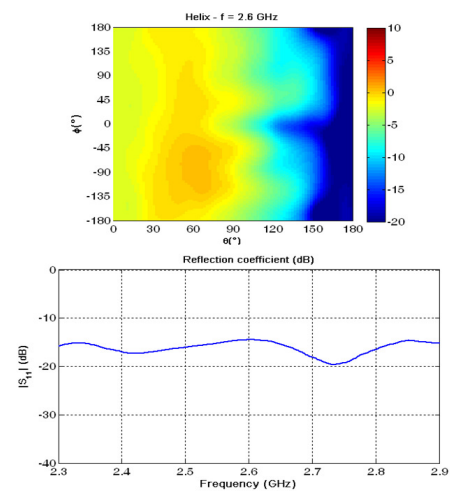

**Figure 2.** Aerial Base Station antenna: Helix antenna measurement results collected during the validation phase.

For the integration of the antennas in the Helikite, i) the antenna orientation depending on polarization and aperture and ii) MIMO functionality are taken into account. A pendulum mount was created and attached to the Helikite in order to ensure that antennas orientation will be vertical whatever the inclination of the balloon. Figure 1(b) gives an insight of the installation recommendations of the LTE sub-systems on the Helikite.

### d. Flying Line and Fiber-optic Cable

The Flying Line is 500 meters long and made of pure Dyneema flying line rated at 1,700 kg breaking strain. As a 34m³ Helikite will not pull more than 600 kg in a 60 mph wind, the breaking strain is acceptable. While, fiber-optic consists of 500m of twin fiber cable (weight is ~1 kg per 100m, diameter equal to 3mm, encapsulated inside a protection).

### B. Terrestrial Segment

The main component of the terrestrial segment is the eNB-BB, which offers a cost effective LTE solution for flexible deployments. It connects to the distributed-EPC for providing a complete end-to-end management solution, which can be deployed in outdoor environments using a baseband cabinet (see Figure 1(c)). The baseband cabinet is equipped with:

- MicroTCA rack aimed at receiving the eNB baseband boards for the PHY and MAC layers,
- Server where the distributed-EPC software and the SIP server software are running,
- Foldable keyboard and screen to have easy access to the server (e.g. to perform registration of new MM-UEs),
- Rack dedicated to routing, cabling and powering functions of the terrestrial segment.

The baseband cabinet is connected to the deployable Ka-band satellite terminal (using Ethernet cable) and to the RRH (using optical fiber). The deployable Ka-band terminal provides satellite backhauling to the AeNB. The demonstrator consists of a portable and fast deployable satellite dish with auto-pointing functionalities and flight-case. ABSOLUTE backhauling framework uses the Eutelsat KA-SAT 9A satellite which offers connectivity over Europe by means of 82 beams through a network of ten ground control stations.

### C. Testing Campaign Results

The main objective of the initial campaign was to test the user equipment attachment to the AeNB, which was flying at 25m above the ground using a transmission power of 23 dBm. Measurements collected for two different types of user equipment are reported in Table 2.

**Table 2.** Measurements at the User Equipment collected during the validation phase of ABSOLUTE demonstration.

| UE Type | Smartphone | Dongle |
|---|---|---|
| **Maximum Reference Signal Received Power (at LAP site)** | -79dBm | -80dBm |
| **Maximum Interference & Noise Ratio (at LAP site)** | N/A | 25dB |
| **Maximum Distance performing ping** | 300m | 562m |
| **Minimum Reference Signal Received Power (maximum distance)** | -100dBm | -110dBm |
| **Minimum Interference & Noise Ratio (maximum distance)** | N/A | 10dB |

## V. AERIAL COMMUNICATION CONSIDERATIONS

Several aerial-terrestrial communication aspects were investigated during the execution of the ABSOLUTE project.

### A. Air-to-Ground Channel Model

In terrestrial communications, the transmitted signal traverses through the urban environment where the RF signal's amplitude decays as a function of the traveled distance. This is usually modeled by a log-distance relation and a path-loss exponent. However, it is observed that radio signal propagation in an Air-to-Ground (A2G) radio channel differs largely from the terrestrial case. This is due to the fact that the radio signals transmitted from an aerial platform propagate through free space until reaching the urban environment where they incur shadowing, scattering and other effects caused by man-made structures. An A2G channel model was developed for low-altitude platforms for different environment conditions using ray tracing simulations [9]. The environment conditions were modeled according to the geometrical statistical parameters given by ITU-R to model high-rise urban, dense urban, urban and suburban areas. It was observed from the results that the A2G path-loss is dependent on the elevation angle given by $\theta$, which is the angle at which the aerial platform is seen from the ground terminals. Figure 3(a) shows the difference between the A2G channel and the terrestrial channel.

The A2G path-loss is modeled with two components. The first component consists of the free space path-loss, while the second part includes the additional path-loss incurred due to the effects caused by the urban or suburban environment, called also as the excessive path-loss. The A2G path-loss can be expressed as follows:

$$PL_\xi = FSPL + \eta_\xi$$

where, $FSPL$ represents the free space path-loss between the aerial platform and the ground terminal and $\xi$ refers to the propagation group divided into two groups: (i) LoS for line-of-sight conditions (good group) and (ii) NLoS for non-line-of-sight conditions (not-so-good group). The excessive path-loss of each propagation group is characterized with different statistical parameters for different environments while the distribution is modeled as Gaussian. On the other hand the probability of a terminal belonging to a certain group, called as group occurrence probability, depends on the elevation angle $\theta$.

### B. Optimal Positioning of Aerial Platform

One of the main advantages of using aerial base-stations is the elevated look-angle which allows covering large areas compared to its terrestrial equivalents. As described above, the A2G channel is composed of FSPL and excessive path-loss $\eta_\xi$ where $\xi$ refers to the propagation group. The probability of a ground terminal falling into line-of-sight group increases with increasing altitude of the aerial platform which also increases the coverage radius of the aerial base-station. Figure 3(b) shows this effect between the aerial base-station's coverage radius and its altitude. However, increasing the aerial platform's altitude increases the distance between the terrestrial ground terminals and the platform. This increases the FSPL component of the path-loss. Therefore, we can observe that there is a trade-off between the FSPL and the excessive path-loss thereby allowing optimization of the aerial platform's altitude to provide maximum coverage. Theoretical optimization of aerial platform's altitude to provide maximum coverage was performed with respect to a maximum allowed path-loss at the ground terminals [9]. Notice that the operational altitude of an aerial platform is constrained by the mechanical properties of the aerial platform itself and aeronautical regulations, which are the main barriers for achieving optimal theoretical altitudes. In the ABSOLUTE demonstration, the altitude of the aerial platform was limited to 150m due to the payload constraints of the Helikite. However, it is expected that the fast evolution of mechanical design of aerial platforms and new aeronautical regulation will allow optimal altitude placement in the future.

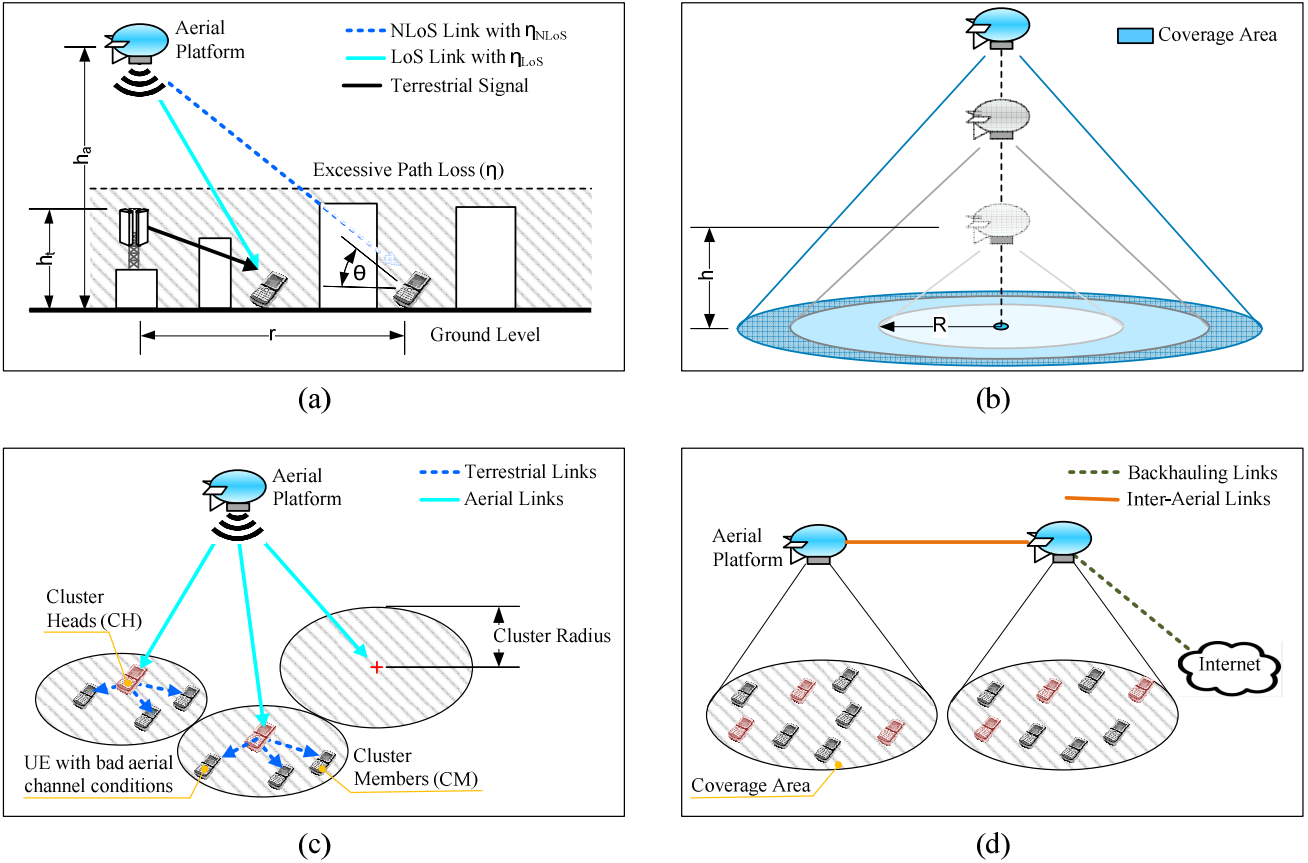

(a)

(b)

(c)

(d)

**Figure 3:** Main consideration taking into account for designing and deploying LTE aerial base stations.

### C. Clustering and Relaying

Since the distance between the terrestrial ground terminals and the aerial platform could be large, there is significant amount of energy required at the ground terminals to communicate with the aerial platform. To provide energy-efficient communications with the aerial platform, a technique called clustering has been investigated. The basic idea behind clustering is to group ground terminals in clusters, so that one node within each partition is responsible of collecting information from other members and forwarding it to the AeNB. Figure 4 shows the energy consumption of the network with and without clustering of nodes in an aerial communication network for different altitudes of aerial platform. We can observe that clustering of ground terminals significantly increases the energy efficiency of the network. Moreover, this approach would significantly reduce the number of terminals attempting to connect with the aerial platform thus easing congestion. Clustering of terrestrial terminals covered by an AeNB is shown in Figure 3(c).

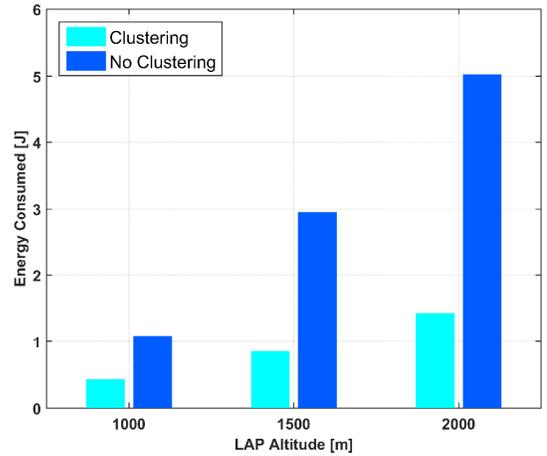

**Figure 4:** Comparison of energy consumption of the terrestrial network with and without clustering of ground terminals.

In harsh environments, some user equipment (UEs) are under bad channel conditions with the aerial base-station prohibiting communications with the aerial platform leaving the UE in outage. Relaying techniques are used in which other nearby UEs relay the information from the uncovered UE to the aerial base-station thus providing coverage [10]. These techniques can also be used to provide coverage extension and capacity improvement to the network.

### D. Wireless Backhauling and Self-organization

To allow for inter-cell and internet connection, the AeNBs connect using wireless backhauling (see Figure 3(d)). Since the aerial network architecture works in dynamic conditions, the aerial base station needs to be endowed with features for self-organization and corresponding cognitive algorithms. The communication between two different aerial base stations is not only to handle handovers, but also to get enough information about the network topology and channel conditions [11]. Satellite and WiFi were considered as candidate technologies for providing wireless backhauling. Satellite backhauling brings the advantage of unlimited coverage offering the possibility of connecting the aerial network for any distance. However, the delay introduced by the satellite links may affect some real time services such as voice and real-time video. To avoid satellite delays and the cost, WiFi links can be used paying the cost of reduced coverage and capacity.

#### a. Frequency Allocations

The cognitive algorithms are the basis to raise awareness about the conditions used for the network set up and to take the best choice for the radio resource management [12]. Several techniques for spectrum allocation have been considered, with dedicated spectrum for aerial base station being a desired solution for avoiding interference. However dynamic spectrum sharing capabilities must also be considered, thus the final demonstration combines:

- **Real-time Data Sensing**: The flexible SDR running in the RRH, was updated in order to implement a cognitive extension with key sensing functionalities for obtaining occupancy thresholds of spectrum and collecting data measurement.
- **Radio Environment Map** is an intelligent database, which stores, processes and delivers information about the status of the radio environment, which is publicly available, over the target area.

Outcomes of these techniques are combined for providing a prioritized list of LTE channels, and sub-channels that are not been utilized so they can be used at each aerial base station over the target area. This is intended for use in future full-scale deployments, on the basis that the ABSOLUTE system will operate an enhanced version of LTE-A, which will dynamically share spectrum with an incumbent LTE-A system, given that dedicated spectrum might not be available.

#### b. Standalone Operations

LTE networks are typically deployed in centralized data centers serving thousands of cells. However, a main requirement within ABSOLUTE network is the ability of the AeNB to operate in standalone manner without relying on centralized equipment. The concept of standalone aerial base stations using LTE technology introduces the necessity of having a distributed EPC embedded at the base station side.

To realize this, the ABSOLUTE project introduced the concept of Flexible Management Entity (FME), which is a software architecture to allow the virtualization and decentralization of the EPC [13]. In fact, FME virtualizes the LTE core network in simplified format which is sufficient for serving a single cell and its subscribers. This software entity is able to run in a small server physically located close to the eNB-BB (as shown in the Figure1(c)). Mobility Management Entity, Serving and Packet Data Network Gateways and Home Subscriber Server as well as their communication interfaces are part of the virtualized EPC architecture supported by FME. Additional units for managing the wireless backhauling operations and routing & topology management protocols implemented for the communication between AeNBs have been embedded as part of FME (for additional details refer to [13]).

### E. Limitations and Remaining Challenges

One limitation of implementing AeNB is that the current telecom equipment are not designed to be placed on aerial platforms as well as aerial platforms are not designed for carrying telecom equipment. So, mechanical limitations, such as maximum payload of the aerial platform or energy sources for powering equipment, have a significant impact on choosing the access technology to be installed on it and the altitude for optimizing the coverage area. The power consumption of the telecom equipment should be carefully considered taking into account that batteries, fuel or solar panels are most likely to be used on aerial platforms [14].

Even if the coverage area of the AeNB is wide, its terrestrial component still needs access to the target area in order to fully utilize its isotropic coverage on the ground. Therefore the placement of the terrestrial segment will influence the flying line of the aerial segment. This is an important limitation of the tethered aerial platform. Additionally, altitude and coverage area of AeNBs also bring issues about regulations in regional borders, military and civil aviation etc. for example impact of the AeNB coverage on neighboring countries. Due to the cost of aerial platforms implementation, limited testing and trial capabilities can limit the amount of preliminary results needed to evaluate a complete system composed of several AeNBs [15]. Finally, in-depth business analysis is required to understand the market potential of LAPs compared to satellite and terrestrial networks.

## VI. CONCLUSIONS

In this paper, we presented the new compelling trend of radio communications consisting of deploying wireless networks using aerial platforms. We first tackled the problem from a general perspective, reviewing existing and upcoming aerial platform solutions. We then narrowed down to the design and development undertaken by the ABSOLUTE project, in which the choice is to use Helikites raised in the sky and that carry battery, antenna and RRH equipment. The Helikite is tethered to the ground and an optical fiber connects the Helikite to the eNB-BB placed on the ground. We showed that Helikites offer a longer enduring, inexpensive and easier to use solution compared to other possible alternatives. Furthermore, we discussed several aspects connected to design and implementation of an aerial network composed of Helikites and LTE-A technology, including the overview of regulatory issues connected to aerial platforms.

The general conclusions we can draw are that regulations and mechanical limitations of the aerial platforms have a strong impact at the moment in deciding the suitable wireless technology to be used in the AeNB as well as the network protocol architecture. However, Helikite enabled aerial platform solutions and LTE-A can be proficiently used to provision Internet access during temporary events and emergencies. These platforms might become even a more stable solution provided that reliability and efficiency of the onboard power system can be enhanced with the possibility of powering equipment over the optical fiber. Nevertheless, ABSOLUTE project made considerable progress addressing several topics on the AeNBs. The optimum technology for inter-aerial platform links connecting the aerial platforms and energy sources for efficiently powering the communication equipment are still an open research issues, which will be part of the future work.

ACKNOWLEDGMENT

The research leading to these results has received partial funding from the EC Seventh Framework Programme (FP7-2011-8) under the Grant Agreement FP7-ICT-318632.

BIOGRAPHIES

Sathyanarayanan Chandrasekharan (s.chandrasekharan.2014@ieee.org) is PhD candidate at the School of Engineering at RMIT University, Melbourne, Australia. He holds a master degree in Network Engineering from RMIT University for which he was selected in the Vice-Chancellor's list of Academic Excellence 2012. He is a recipient of the Australian Post Graduate Award funded by the Australian Government and the Orange Labs scholarship.

Karina Gomez Chavez (karina.gomezchavez@rmit.edu.au) received her Master degree in Wireless Systems and Related Technologies from the Turin Polytechnic in 2006, Italy. In 2007, she joined Communication and Location Technologies Area at FIAT Research Centre. In 2008, she joined Future Networks Area at Create-Net, Italy. In 2013, she obtained her PhD degree in Telecommunications from the University of Trento, Italy. Since July 2015, she is lecturer at School of Engineering in RMIT University, Melbourne, Australia.

Akram Al-Hourani is PhD candidate at the School of Engineering at RMIT University, Melbourne, Australia, where he is a recipient of the Australian Post Graduate Award funded by the Australian government, he has also been a recipient of Orange Labs scholarship. Akram has worked for 7 years as a radio network planning engineer for mobile telecom industry, and then for as an ICT project manager for several projects spanning over different wireless technologies.

Sithamparanathan Kandeepan has a PhD from the University of Technology, Sydney and is currently with the School of Engineering at the RMIT University. He had worked with the NICTA Research Laboratory (Canberra) and CREATE-NET Research Centre (Italy). He is a Senior Member of the IEEE, had served as a Vice Chair of the IEEE Technical Committee on Cognitive Networks, and currently serves as the Chair of IEEE VIC region Communications Society Chapter.

**Tinku Rasheed** (tinku.rasheed@create-net.org) is a Senior Research staff member at Create-Net. Since May 2013, He is heading the Future Networks R&D Area [FuN] within Create-Net. Dr. Rasheed has extensive industrial and academic research experience in the areas of mobile wireless communication and data technologies, end-to-end network architectures and services. He has several granted patents and has published his research in major journals and conferences.

**Leonardo Goratti** (leonardo.goratti@create-net.org) received his PhD degree in Wireless Communications in 2011 from the University of Oulu-Finland and his M.Sc. in Telecommunications engineering in 2002 from the University of Firenze-Italy. From 2003 until 2010, He worked at the Centre for Wireless Communications Oulu-Finland. From 2010 until early 2013 He worked at the European funded Joint Research Centre (JRC) of Ispra, Italy. In 2013, He joined the Research Centre CREATE-NET Trento-Italy.

**Laurent Reynaud** (laurent.reynaud@orange.com) is a senior research engineer for the Future Networks research community at Orange. After receiving his engineering degree from ESIGETEL at Fontainebleau in 1996, He acquired a experience regarding the development and deployment of distributed software in the context of telecommunications, through successive positions in the French Home Department in 1997, in Alcatel-Lucent from 1998 to 2000, and in Orange since 2000. He participated to many French, European and international research projects.

**David Grace** (david.grace@york.ac.uk) received his PhD from University of York in 1999. Since 1994 He has been in the Department of Electronics at York, where He is now Professor (Research) and Head of Communications and Signal Processing Group. Current research interests include aerial platform based communications, cognitive dynamic spectrum access and interference management. He is currently a Non-Executive Director of a technology start-up company, and a former chair of IEEE Technical Committee on Cognitive Networks.

**Isabelle Bucaille** (isabelle.bucaille@thalesgroup.com) received the engineering degree from ISEP in France in 1994. Then she joined the CNI Division of TH-CSF for digital processing studies. She participated in 1997 to the ETSI group in charge of HiperLAN2 normalisation. In 1998 she was in charge of system definition concerning Stratospheric Platforms (HAPS). Since September 2001 she is in the Secured Wireless Products Activity of THALES Communications. Since 2011, she has been appointed TCS representative in 3GPP.

**Thomas Wirth** (thomas.wirth@hhi.fraunhofer.de) received a Dipl.-Inform. degree in computer science from the Universität Würzburg, Germany, in 2004. In 2004, He joined Universität Bremen working in the field of robotics. In 2006 He joined HHI working as senior researcher on resource allocation algorithms for LTE/LTE-Advanced systems. Since 2011, Thomas is head of the Software Defined Radio (SDR) group in the Wireless Communications and Networks Department, working in various projects on PHY and MAC design for 5G.

**Sandy Allsopp** is the Managing Director and joint owner of Allsopp Helikites Ltd. He is designer of the Helikite aerostat in 1993 and holder of Helikite patents and designs. He is highly experienced in all aspects of Helikite aerostat manufacturer and operations. He has also long experience of many major radio-relay trials and operations.